\documentclass[12pt]{article}
\usepackage{graphicx,amssymb,amsmath,empheq,amsthm}
\usepackage{authblk}
\usepackage{comment}
\usepackage{braket}
\usepackage{subfig} 
\usepackage{soul}
\usepackage{hyperref}
\hypersetup{colorlinks = true, linkcolor=black, citecolor=black, urlcolor=blue}
\usepackage{url}
\usepackage{enumitem}
\usepackage[nosort]{cite}
\usepackage{amsmath,bm}

\theoremstyle{plain}

\usepackage{tikz}
\usepackage{tikz-cd}
\usetikzlibrary{arrows}
\usetikzlibrary{intersections}
\usetikzlibrary{shapes.geometric}
\usetikzlibrary{decorations.pathmorphing, patterns,shapes}
\usetikzlibrary{decorations.markings}

\tikzset{
	mid arrow/.style={postaction={decorate,decoration={
				markings,
				mark=at position .575 with {\arrow{stealth}}
	}}},
	near arrow/.style={postaction={decorate,decoration={
				markings,
				mark=at position .275 with {\arrow{stealth}}
	}}},
	far arrow/.style={postaction={decorate,decoration={
				markings,
				mark=at position .800 with {\arrow{stealth}}
	}}},
	snake arrow/.style={fixed point arithmetic, decorate, decoration={snake,amplitude=2pt, segment length=11pt},postaction={decoration={markings,mark=at position 0.625 with {\arrow{stealth}}},decorate}},
}

\tikzset{
  baseline = -0.5ex,
  wavy/.style = {
    thick,
    decorate,
    decoration={snake,amplitude=2pt,segment length=5pt}},
  swavy/.style = {
    thick,
    decorate,
    decoration={snake,amplitude=1.2pt,segment length=3pt}},
  sdot/.style = {
    circle,
    draw=none,
    fill=black,
    minimum size=2.5pt,
    inner sep=0pt},
  bdot/.style = {
    circle,
    draw=none,
    fill=black,
    minimum size=4pt,
    inner sep=0pt},
  svertex/.style = {
    circle,
    draw=black,
    thick,
    fill=lightgray,
    minimum size=8pt,
    inner sep=1pt},
  mmvertex/.style = {
    circle,
    draw=black,
    thick,
    fill=lightgray,
    minimum size=10pt,
    inner sep=1pt},
  mvertex/.style = {
    circle,
    draw=black,
    thick,
    fill=lightgray,
    minimum size=12pt,
    inner sep=1pt},
  bvertex/.style = {
    circle,
    draw=black,
    thick,
    fill=lightgray,
    minimum size=16pt},
  bbvertex/.style = {
    circle,
    draw=black,
    thick,
    fill=lightgray,
    minimum size=20pt}
 }

\title{A Zoo of Fermionic Relative Rényi Negativities\\ in the Sachdev–Ye–Kitaev Models }

\author[1]{Linjing Jiang}
\author[1,2]{Pengfei Zhang\thanks{PengfeiZhang.physics@gmail.com}}

\affil[1]{\normalsize \it State Key Laboratory of Surface Physics \& Department of Physics \& Center for Field Theory and Particle Physics, Fudan University, Shanghai, 200438, China}
\affil[2]{\normalsize \it Hefei National Laboratory, Hefei 230088, China}

\date{\today}

\begin{document}
\maketitle

\begin{abstract}
Quantum entanglement is a fundamental resource that gives rise to correlations with no classical counterpart in strongly correlated fermionic systems. However, its characterization in mixed quantum states remains challenging. Established entanglement diagnostics, such as logarithmic negativity, are computationally demanding in interacting systems, and their evaluation by exact diagonalization is restricted to small system sizes. Meanwhile, conventional R\'enyi proxies, including twisted and untwisted R\'enyi negativities, can be positive even for separable states. In this work, we introduce a family of \textit{relative R\'enyi negativities} that serve as entanglement witnesses. By relating fermionic partial time reversal to the size superoperator, we establish a path-integral representation for these R\'enyi negativities, enabling their evaluation in thermal states of the Sachdev-Ye-Kitaev (SYK) model and its generalizations via a large-$N$ expansion. Applying this method, we investigate entanglement in both the original SYK model and the Maldacena-Qi coupled SYK model. We identify a special class of relative R\'enyi negativities which serve as plausible proxies for logarithmic negativity and uncover first-order entanglement transitions that are invisible to simple observables. These results highlight the utility of relative R\'enyi negativities as probes of mixed-state entanglement in strongly correlated fermionic systems and suggest the existence of a new class of transitions where these negativities serve as free energies.
\end{abstract}

\bigskip

\section{Introduction}

Quantum entanglement offers an intriguing perspective on novel phenomena in quantum many-body systems. For pure quantum states, bipartite entanglement is commonly characterized by the von Neumann entanglement entropy $S_A=-\operatorname{tr}[\rho_A\ln\rho_A]$, where $\rho_A$ is the reduced density matrix of a subregion $A$. This entropy vanishes when the total state factorizes between $A$ and its complement. Theoretical studies have revealed numerous connections between entanglement entropy and fundamental concepts in many-body physics: For example, in two-dimensional gapped systems with topological order, the entanglement entropy contains a universal subleading contribution, known as the topological entanglement entropy, that encodes global topological properties of the many-body wave function~\cite{PhysRevLett.96.110404,PhysRevLett.96.110405}. The dynamics of entanglement entropy following a quantum quench reveal the buildup of quantum entanglement and probe the rate at which quantum information spreads, distinguishing thermalizing systems from localized ones~\cite{Calabrese:2009qy,PhysRevLett.111.127205,PhysRevLett.110.260601,Huse:2014tqa,Nandkishore:2014kca}. Recent studies of measurement-induced phase transitions have further shown that the scaling of entanglement entropy can serve as an order parameter for transitions driven by competition between unitary evolution and projective measurements~\cite{Li:2018mcv,Skinner:2018tjl,Chan:2018upn}. On the experimental side, direct measurement of the von Neumann entropy in many-body systems remains challenging. Its one-parameter generalization, the R\'enyi entropy $S_A^{(n)}=\frac{1}{1-n}\ln\operatorname{tr}(\rho_A^n)$, offers an experimentally accessible alternative and has been measured using randomized measurement protocols~\cite{Brydges:2019wut} or interference between multiple copies of the system~\cite{Islam:2015mom}. In addition, R\'enyi entropies are well suited to large-scale numerical simulations~\cite{Zhao2021MeasuringRE} and path-integral analysis in solvable models~\cite{Zhang:2020kia,zhang2022quantum}.

By contrast, for mixed quantum states, the subsystem von Neumann entropy contains both thermal and entanglement contributions and is therefore not, by itself, a measure of quantum entanglement. Instead, logarithmic entanglement negativity is widely used to diagnose mixed-state entanglement~\cite{Audenaert:2002xfl,Ferraro:2008zz,Cavalcanti:2008kzj,Marcovitch:2008sxc,Wichterich:2008vfx,Ruggiero:2016aqr,Wichterich:2010btp,Santos:2011rww,Castelnovo:2013ozs,Lee:2013tth,Wen:2016snr,Wen:2016bla,Calabrese:2012ew,Calabrese:2012nk,Calabrese:2014yza,Blondeau-Fournier:2015yoa,Eisler:2014dze,Sherman:2015bjw,Coser:2014gsa,
Hoogeveen:2014bqa,Wen:2015qwa,Eisler:2015tgq,Eisler:2016bfd,Chang2016EntanglementNI,Eisert:2016xgf,Shapourian:2018ozl,Shapourian:2018lsz,Lu:2018yxh,Lu:2019xwg,Lu:2019owx,Choi:2023nrx,Parez:2023xpj,Fang:2025nyw,Wang:2024fbt}. It is an entanglement monotone whose nonzero value certifies quantum entanglement~\cite{PhysRevA.58.883,PhysRevLett.77.1413,Horodecki:1996nc,Eisert:1998pz,PhysRevA.65.032314,PhysRevLett.95.090503}. The logarithmic entanglement negativity is defined as $\mathcal{N}_A=\ln\|\rho^{T_A}\|_1$, where $\rho^{T_A}$ denotes the partial transpose of the density matrix $\rho$ with respect to subsystem $A$ and $\|O\|_1=\operatorname{tr}\sqrt{O^\dagger O}$ is the trace norm. For bosonic systems, this operation is defined by $\bigl[(|i\rangle\langle k|)_A\otimes(|j\rangle\langle l|)_B\bigr]^{T_A}=(|k\rangle\langle i|)_A\otimes(|j\rangle\langle l|)_B$ and extended linearly to arbitrary operators. For fermionic systems, additional structures arise from the existence of several definitions of the fermionic partial transpose~\cite{Eisler:2015tgq,Shapourian:2016cqu,Shapourian:2019xfi}. By analogy with the relation between R\'enyi entropies and the von Neumann entropy, the R\'enyi negativity ratios $\mathcal{R}^{(n)}_A=\frac{1}{1-n}\ln\left[\frac{\operatorname{tr}[(\rho^{T_A})^n]}{\operatorname{tr}(\rho^n)}\right]$ have been introduced as computationally accessible proxies for logarithmic negativity, enabling large-scale numerical simulations~\cite{Calabrese:2012ew,Calabrese:2012nk,Calabrese:2013mi,Alba:2013mg,Chung:2014pwd,DeNobili:2015dla,Coser:2015mta,Coser:2015eba,Coser:2015dvp,Cornfeld:2018sac,Gray:2018fbo,Elben:2020hpu,Neven:2021igr,Yu:2021uzg,Feldman:2022hrf,Wu:2019qvm,Murciano:2021zvs,Wang:2023kdg,Ding:2025upg,Wang:2025mdu,Liu2026EntanglementRN}. Nevertheless, these quantities can be nonzero even for separable states and thus do not serve as entanglement witnesses.

In this work, we introduce a zoo of R\'enyi negativities for fermionic many-body systems by exploiting the rich structure of the fermionic partial transpose. By combining the partially transposed density matrix and its Hermitian conjugate in different orders, we construct a family of distinct probes of mixed-state entanglement. We show that these quantities vanish for separable states, so a nonzero value of any member certifies entanglement. By relating fermionic partial time reversal to the size superoperator~\cite{roberts2018operator,Qi:2018bje,Zhang:2022fma,Liu:2023lyu}, we derive a path-integral representation at fixed integer replica number, without requiring a replica limit. This formulation provides a framework for both analytical calculations and large-scale numerical studies. To investigate the behavior of these quantities in a controlled setting, we develop a large-$N$ expansion and apply it to both the original Sachdev-Ye-Kitaev (SYK) model~\cite{maldacena2016remarks,kitaev2015simple,kitaev2018soft} and the Maldacena-Qi coupled SYK model~\cite{Maldacena:2018lmt}. Our analysis identifies a special class of relative R\'enyi negativities which serve as plausible proxies for logarithmic negativity and reveals first-order entanglement transitions that are invisible to simple observables. These findings highlight the utility of relative R\'enyi negativities as probes of mixed-state entanglement in strongly correlated fermionic systems and suggest a new class of transitions for which these negativities serve as free energies.

\section{A Zoo of Relative R\'enyi Negativities }
We are interested in characterizing mixed-state entanglement in fermionic many-body systems. Without loss of generality, we consider a system of $N$ Majorana fermions $\chi_j$, with $j\in\{1,2,\ldots,N\}$. Following the convention commonly used in the SYK literature~\cite{maldacena2016remarks}, the canonical anticommutation relations are $\{\chi_j,\chi_k\}=\delta_{jk}$. In fermionic systems, the definition of the entanglement negativity for a subsystem $A$ relies on the choice of the fermionic partial transpose operation. In their early work, Eisler and Zimbor\'as~\cite{Eisler:2015tgq} defined the fermionic partial transpose through the ordinary partial transpose of the corresponding spin density matrix obtained via a Jordan-Wigner transformation~\cite{altland2010condensed}. Here, we adopt an alternative perspective by specifying its action on Pauli strings~\cite{Shapourian:2016cqu}. We begin by expanding a generic density matrix $\rho$ in a basis of Majorana string operators:
\begin{equation}\label{eq:rho}
\rho=\sum_{\bm{a},\bm{b}}c_{\bm{a},\bm{b}}(\rho)\,
i^{s(s-1)/2}2^{s/2}
\chi_1^{a_1}\chi_2^{a_2}\cdots\chi_{N_A}^{a_{N_A}}
\chi_{N_A+1}^{b_1}\cdots\chi_N^{b_{N_B}}.
\end{equation}
Here, subsystem $A$ contains the first $N_A$ Majorana fermions, with $N_A$ even, while subsystem $B$ contains the remaining $N_B=N-N_A$ Majorana fermions. The binary vectors $\bm{a}\in\{0,1\}^{N_A}$ and $\bm{b}\in\{0,1\}^{N_B}$ specify which Majorana operators appear in each string, while the expansion coefficients $c_{\bm{a},\bm{b}}(\rho)$ are real. The string length $s=|\bm{a}|+|\bm{b}|$, where $|\bm{a}|$ and $|\bm{b}|$ count the nonzero entries in the respective binary vectors, is referred to as the operator size~\cite{roberts2018operator,Qi:2018bje,Zhang:2022fma,Liu:2023lyu}. Accordingly, $|\bm{a}|$ gives the operator size within subsystem $A$~\cite{Zhang:2023xrr}. The prefactor $i^{s(s-1)/2}2^{s/2}$ ensures that each basis string is Hermitian and squares to the identity. Then, the partial transpose introduced by Eisler and Zimbor\'as~\cite{Eisler:2015tgq} is represented by the linear map $\rho^{\mathcal{T}_A} \equiv \mathcal{T}_A[\rho]$ with:
\begin{equation}
 \mathcal{T}_A[\chi_1^{a_1}\cdots\chi_{N_A}^{a_{N_A}}
\chi_{N_A+1}^{b_1}\cdots\chi_N^{b_{N_B}}]=(-1)^{|\bm{a}|(|\bm{a}|-1)/2}\chi_1^{a_1}\cdots\chi_{N_A}^{a_{N_A}}
\chi_{N_A+1}^{b_1}\cdots\chi_N^{b_{N_B}}.
\end{equation}
Nevertheless, this definition has several undesirable features. For example, it generally fails to satisfy a natural identity $(\rho^{\mathcal{T}_A})^{\mathcal{T}_{\bar{A}}}=\rho^\mathcal{T}$ because the quadratic phase exponent $|\bm{a}|(|\bm{a}|-1)/2$ is not additive. Moreover, it can map Gaussian states to non-Gaussian states, introducing additional complications even in free-fermion systems\footnote{This becomes evident when the operation is expressed in terms of the size superoperator, as we show below.}.

To address these issues, Shapourian et al.~\cite{Shapourian:2016cqu} introduced an alternative definition of the fermionic partial transpose, also known as partial time reversal. We denote this linear map by $T_A$ and write $\rho^{T_A}=T_A[\rho]$. This map satisfies:
\begin{equation}\label{eq:partial_transpose}
 T_A[\chi_1^{a_1}\cdots\chi_{N_A}^{a_{N_A}}
\chi_{N_A+1}^{b_1}\cdots\chi_N^{b_{N_B}}]=i^{|\bm{a}|}\chi_1^{a_1}\cdots\chi_{N_A}^{a_{N_A}}
\chi_{N_A+1}^{b_1}\cdots\chi_N^{b_{N_B}}.
\end{equation}
The phase exponent is now linear in the subsystem operator size $|\bm{a}|$, so the map satisfies $(\rho^{T_A})^{T_{\bar{A}}}=\rho^{T}$. When $A$ is the entire system, $s=|\bm{a}|$, and only even-$s$ terms contribute to a density matrix that preserves fermion parity. The transformation then replaces $i^{s(s-1)/2}$ with $(-i)^{s(s-1)/2}$ in the expansion \eqref{eq:rho}. Its action on such density matrices is therefore equivalent to a time-reversal transformation that sends $i\rightarrow -i$ while leaving the Majorana operators unchanged, $\chi_j\rightarrow\chi_j$. Adopting this convention, the logarithmic negativity $\mathcal{N}_A$ and the (untwisted) R\'enyi negativity $\mathcal{N}_A^{(2n)}$ are defined as
\begin{equation}
\begin{aligned}
\mathcal{N}_A
=\ln\|\rho^{T_A}\|_1=\ln\operatorname{tr}\sqrt{(\rho^{T_A})^\dagger\rho^{T_A}},\ \ \ \ \ \ 
\mathcal{N}_A^{(2n)}
=\frac{1}{1-2n}\ln\operatorname{tr}\left[(\rho^{T_A})^{2n}\right].
\end{aligned}
\end{equation}
In bosonic systems, the logarithmic negativity can be recovered by analytically continuing the even-order R\'enyi negativities, $\mathcal{N}_A=\lim_{n\rightarrow 1/2}(1-2n)\mathcal{N}_A^{(2n)}$. In fermionic systems, however, the partially transposed density matrix is generally non-Hermitian, $(\rho^{T_A})^\dagger\neq\rho^{T_A}$, so the same analytic continuation of the untwisted R\'enyi negativities does not generally yield the logarithmic negativity. Moreover, $\mathcal{N}_A^{(n)}$ can be complex for certain states. These properties motivate the introduction of twisted R\'enyi negativities for fermionic systems~\cite{Shapourian:2019xfi}:
\begin{equation}
\tilde{\mathcal{N}}_A^{(2n)}
=\frac{1}{1-2n}\ln\operatorname{tr}\left[\left((\rho^{T_A})^\dagger\rho^{T_A}\right)^{n}\right].
\end{equation}
The logarithmic negativity is recovered through $\mathcal{N}_A=\lim_{n\rightarrow 1/2}(1-2n)\tilde{\mathcal{N}}_A^{(2n)}$. These R\'enyi negativities are by definition real and non-negative. We further introduce the untwisted/twisted R\'enyi negativity ratios, $$\mathcal{R}_A^{(2n)}=\mathcal{N}_A^{(2n)}-S^{(2n)}\ \ \ \ \ \tilde{\mathcal{R}}_A^{(2n)}=\tilde{\mathcal{N}}_A^{(2n)}-S^{(2n)},$$ by subtracting the thermal R\'enyi entropy $S^{(2n)}=\frac{1}{1-2n}\ln\operatorname{tr}[\rho^{2n}]$. In particular, the twisted ratio $\tilde{\mathcal{R}}_A^{(2)}$ vanishes for any density matrix $\rho$, because the partial transpose phase factors in $(\rho^{T_A})^\dagger$ and $\rho^{T_A}$ cancel in the trace. Thus, the fourth-order ratio $\tilde{\mathcal{R}}_A^{(4)}$ is the first nontrivial member of this family. In the lattice models studied in Ref.~\cite{Wang:2025mdu}, this quantity exhibits desirable properties as an entanglement diagnostic, including a monotonic decrease with increasing temperature. However, subtracting the thermal entropy does not guarantee that a positive value of $\mathcal{R}_A^{(2n)}$ or $\tilde{\mathcal{R}}_A^{(2n)}$ certifies entanglement, and these quantities can be positive even for simple separable states (see Appendix~\ref{example1} for an example with $N=8$). 

We resolve this problem by introducing an alternative family of R\'enyi negativities whose members serve as entanglement witnesses for mixed states. To motivate their definition, we recall that separable states in fermionic systems can be written as
\begin{equation}
\rho_{\mathrm{sep}}=\sum_\alpha p_\alpha\,\rho_{A,\alpha}\otimes\rho_{B,\alpha},
\end{equation}
where $\rho_{A,\alpha}$ and $\rho_{B,\alpha}$ are local density matrices that conserve fermion parity, and $p_\alpha$ are the corresponding probabilities. In particular, $\rho_{A,\alpha}$ contains only Majorana strings with even $|\bm{a}|$. Consequently, applying the superoperator $T_A$ introduces only real phase factors $\pm1$, so the partially transposed separable state remains Hermitian:
\begin{equation}
\rho^{T_A}_{\mathrm{sep}}=\sum_\alpha p_\alpha\,(\rho_{A,\alpha}^{T_A})\otimes\rho_{B,\alpha}=\sum_\alpha p_\alpha\,(\rho_{A,\alpha}^{T_A})^\dagger\otimes\rho_{B,\alpha}^\dagger=(\rho^{T_A}_{\mathrm{sep}})^\dagger.
\end{equation}
Thus, any quantity that detects a difference between $\rho^{T_A}$ and $(\rho^{T_A})^\dagger$ can serve as an entanglement witness in fermionic systems. Motivated by this observation, we introduce a class of generalized R\'enyi moments by combining $\rho^{T_A}$ and $(\rho^{T_A})^\dagger$ in different orderings:
\begin{equation}
\mathcal{M}^{\bm{f}}_A=\operatorname{tr}\left[X_{f_1}X_{f_2}\cdots X_{f_{2n}}\right],\ \ \ \ \ \ \bm{f}=(f_1,f_2,\cdots,f_{2n}).
\end{equation}
Here, each entry of $\bm{f}$ satisfies $f_j\in\{+,-\}$, and the corresponding operators are defined as $X_+=\rho^{T_A}$ and $X_-=(\rho^{T_A})^\dagger$. Naively, there are $2^{2n}$ possible moments at R\'enyi order $2n$. However, the number of independent moments is significantly reduced by two symmetries: (1) $\mathcal{M}_A^{\bm{f}}$ is invariant under cyclic permutations of $\bm{f}$, owing to the cyclicity of the trace; (2) $\mathcal{M}_A^{\bm{f}}=\mathcal{M}_A^{-\bm{f}}$, where $-\bm{f}$ denotes the simultaneous reversal of all signs, because $(\rho^{T_A})^\dagger=P_A\rho^{T_A}P_A$, with $P_A$ the fermion-parity operator of subsystem $A$. Therefore, in the following discussion, we choose only one representative $\bm{f}$ from each equivalence class. For example, at second order, we have $\bm{f}\in\{(+,+),\ (+,-)\}$, while at fourth order, representatives of the four classes are
\begin{equation}
\bm{f}\in\{(+,+,+,+),\ (+,+,+,-),\ (+,+,-,-),\ (+,-,+,-)\}.
\end{equation}
In this notation, the moments appearing in the untwisted and twisted R\'enyi negativities correspond to $\bm{f}_u=(+,+,+,+,\ldots)$ and $\bm{f}_t=(+,-,+,-,\ldots)$, respectively.

For separable states, all moments at the same R\'enyi order $2n$ are identical. Therefore, a nonzero difference between moments of the same order witnesses entanglement in the mixed state $\rho$. We choose the moment corresponding to the alternating sign sequence $\bm{f}_t$ as a reference and introduce a zoo of \textit{relative R\'enyi negativities} as
\begin{equation}
\mathcal{P}^{\bm{f}}_A\equiv -\ln \left(\left|\mathcal{M}^{\bm{f}}_A\right|/ \mathcal{M}^{\bm{f}_t}_A\right)=-\ln\left[\frac{|\operatorname{tr}\left[X_{f_1}X_{f_2}\cdots X_{f_{2n}}\right]|}{\operatorname{tr}\left[\left((\rho^{T_A})^\dagger\rho^{T_A}\right)^{n}\right]}\right]. 
\end{equation}
By definition, $\mathcal{P}^{\bm{f}}_A$ is real for arbitrary $\bm{f}$. Moreover, it is non-negative by applying the generalized H\"older inequality~\cite{Watrous:2018zil}:
\begin{equation}
\begin{aligned}
\left|\operatorname{tr}\left[
X_{f_1}X_{f_2}\cdots X_{f_{2n}}
\right]\right|
&\leq \prod_{j=1}^{2n}\|X_{f_j}\|_{2n}=(\|\rho^{T_A}\|_{2n})^{2n}=\operatorname{tr}\left[
\left((\rho^{T_A})^\dagger\rho^{T_A}\right)^n
\right],
\end{aligned}
\end{equation}
where $\|O\|_p=\left[\operatorname{tr}[(O^\dagger O)^{p/2}]\right]^{1/p}=\|O^\dagger\|_p$ is the Schatten $p$-norm. This guarantees that $\mathcal{P}^{\bm{f}}_A\geq0$ and $\mathcal{P}^{\bm{f}_t}_A=0$. One can also show that the relative R\'enyi negativities satisfy the condition $\mathcal{P}^{\bm{f}}_A=\mathcal{P}^{\bm{f}}_B$ for arbitrary $\bm{f}$, as demonstrated in Appendix~\ref{proof}. We further highlight two special choices of $\bm{f}$: 
\begin{enumerate}
\item  If we take $\bm{f}=\bm{f}_u$, the resulting relative R\'enyi negativity $\mathcal{P}^{\bm{f}_u}_A$ is related to the difference between the twisted and untwisted R\'enyi negativities:
\begin{equation}
\mathcal{P}^{\bm{f}_u}_A=(2n-1)\left(\text{Re}[\mathcal{N}_A^{(2n)}]-\tilde{\mathcal{N}}_A^{(2n)}\right)=(2n-1)\left(\text{Re}[\mathcal{R}_A^{(2n)}]-\tilde{\mathcal{R}}_A^{(2n)}\right).
\end{equation}
In particular, we have $\mathcal{P}^{(+,+)}_A=\operatorname{Re}[\mathcal{R}_A^{(2)}]$ and $\mathcal{P}^{(+,+,+,+)}_A=3\left(\operatorname{Re}[\mathcal{R}_A^{(4)}]-\tilde{\mathcal{R}}_A^{(4)}\right)$. The logarithmic negativity can be obtained from $\mathcal{P}^{\bm{f}_u}_A$ by analytic continuation to $n=1/2$. 

\item We choose $\bm{f}=\bm{f}_m$ as a minimal deviation from the alternating sign sequence $\bm{f}_t$, obtained by flipping a single entry. All such choices are equivalent under cyclic permutations and simultaneous sign reversal. At second order, this choice coincides with $\bm{f}_u$, while at fourth order, we may choose $\bm{f}_m=(+,+,+,-)$.

\end{enumerate}

Although positive values of the relative R\'enyi negativities certify quantum entanglement, this does not guarantee that they exhibit desirable behavior as physical parameters, such as temperature, vary. Understanding this dependence is essential if their magnitudes are to be interpreted as indicators of the strength of entanglement. This motivates a systematic investigation of these quantities in concrete solvable models, as we discuss below.

\section{Negativities in the SYK Model}

We now calculate the relative R\'enyi negativities in the Sachdev-Ye-Kitaev model, which describes Majorana fermions with random $q$-body interactions. The Hamiltonian reads~\cite{maldacena2016remarks,kitaev2015simple,kitaev2018soft}
\begin{equation}
H=\sum_{1\leq i_1<i_2<\cdots<i_q\leq N}J_{i_1i_2\cdots i_q}~i^{\frac{q(q-1)}{2}}\chi_{i_1}\chi_{i_2}\cdots\chi_{i_q}
\end{equation}
Here, $J_{i_1i_2\cdots i_q}$ are random couplings drawn from independent Gaussian distributions, with mean and variance given by
\begin{equation}
\overline{J_{i_1i_2\cdots i_q}}=0,\ \ \ \ \ \ \overline{J_{i_1i_2\cdots i_q}^2}=\frac{(q-1)!J^2}{N^{q-1}}.
\end{equation}
For $q=2$, the model describes free fermions with random hopping amplitudes, whereas even $q\geq4$ gives rise to an interacting, chaotic system with all-to-all couplings. The model admits a controlled large-$N$ expansion, yielding saddle-point equations that become exact in the $N\to\infty$ limit and determine its thermodynamic properties~\cite{maldacena2016remarks}. Subsequent studies have extended this method to investigate information-theoretic quantities, including the second R\'enyi entropy~\cite{Zhang:2020kia}, stabilizer R\'enyi entropy~\cite{Zhang:2025rky,Sun:2026ewq,Bettaque:2026vpl}, participation entropy~\cite{Liu:2026yeo}, and fermionic non-Gaussianity~\cite{Garcia-Garcia:2026fpf,Sun:2026qsk,Hoshino:2026fsk}. These studies have revealed new phases and phase transitions that are not captured by conventional thermodynamic observables.

\subsection{Path-Integral Formalism}
We develop a path-integral framework for analyzing negativities in thermal states. For concreteness, we focus on the SYK model, although the path-integral construction applies to general Majorana models before a specific Hamiltonian is introduced. The thermal density matrix is $\rho=e^{-\beta H}/Z$, where $Z=\operatorname{tr}[e^{-\beta H}]$ is the thermal partition function. As discussed in the previous section, subsystem $A$ contains the first $N_A$ Majorana fermions. To construct the partially transposed density matrix $\rho^{T_A}$, we note that the transformation $T_A$, defined in Eq.~\eqref{eq:partial_transpose}, can be expressed in terms of the (subsystem) size superoperator $S_A$:
\begin{equation}
T_A=e^{i\frac{\pi}{2}S_A},\ \ \ \ \ \ S_A[\chi_1^{a_1}\cdots\chi_{N_A}^{a_{N_A}}
\chi_{N_A+1}^{b_1}\cdots\chi_N^{b_{N_B}}]=|\bm{a}|~\chi_1^{a_1}\cdots\chi_{N_A}^{a_{N_A}}
\chi_{N_A+1}^{b_1}\cdots\chi_N^{b_{N_B}}.
\end{equation}
An explicit form of the size superoperator has been identified in Refs.~\cite{roberts2018operator,Qi:2018bje,Zhang:2022fma} and is quadratic in Majorana fermion operators acting on both sides of the density operator:
\begin{equation}
S_A[O]=\sum_{j\in A}\left(\frac{O}{2}-\chi_j O\chi_j\right)
\equiv\sum_{j\in A}\left(\frac{1}{2}-\chi_j\cdot\chi_j\right)[O].
\end{equation}
Here, we assume that $O$ is bosonic, and the notation $A\cdot B$ denotes the superoperator $(A\cdot B)[O]=AOB$. For fermionic operators, additional phases arise from the anticommutation relations between $O$ and $\chi_j$~\cite{gu2022two}. Thus, $T_A=\exp[i\frac{\pi}{2}\sum_{j\in A}(\frac{1}{2}-\chi_j\cdot\chi_j)]$ is a Gaussian superoperator. Similarly, $(\rho^{T_A})^\dagger$ is obtained by applying the Gaussian superoperator $T_A^{-1}=\exp[-i\frac{\pi}{2}\sum_{j\in A}(\frac{1}{2}-\chi_j\cdot\chi_j)]$ to the density matrix $\rho$. In contrast, the phase exponent of $\mathcal{T}_A$ is quadratic in $|\bm{a}|$, so this operation does not correspond to a Gaussian superoperator.

This identification facilitates the development of a simple path-integral representation for the R\'enyi negativities. Our goal is to express the generalized R\'enyi moments $\mathcal{M}^{\bm{f}}_A$ as functional integrals. We begin with a pictorial representation of the partially transposed density matrix $\rho^{T_A}$ and its Hermitian conjugate $(\rho^{T_A})^\dagger$:
\begin{equation}
\rho^{T_A}=\frac{1}{Z}\times
\begin{tikzpicture}[baseline={([yshift=0pt]current bounding box.center)}]
   \draw (1,-0.2) arc(0:-180:1 and 1);
   \draw (1.3,-0.2) arc(0:-180:1.3 and 1.3);
   \draw[dotted,thick] (0.866,-0.725) -- (1.126,-0.875);
   \draw[dotted,thick] (0.5,-1.091) -- (0.65,-1.35);
   \draw[dotted,thick] (0,-1.225) -- (0,-1.525);
   \draw[dotted,thick] (-0.5,-1.091) -- (-0.65,-1.35);
   \draw[dotted,thick] (-0.866,-0.725) -- (-1.126,-0.875);
   \draw[dashed,thick,red] (1,-0.2) -- (-1,-0.2);
   \draw (0.4,-0.8) node{$A$};
   \draw (0.9,-1.4) node{$B$};
   \draw (-1.5,-0.2) node{$\beta$};
   \draw (1.5,-0.2) node{$0$};
\end{tikzpicture},\ \ \ \ \ \ \ \ 
(\rho^{T_A})^\dagger=\frac{1}{Z}\times
\begin{tikzpicture}[baseline={([yshift=0pt]current bounding box.center)}]
   \draw (1,-0.2) arc(0:-180:1 and 1);
   \draw (1.3,-0.2) arc(0:-180:1.3 and 1.3);
   \draw[dotted,thick] (0.866,-0.725) -- (1.126,-0.875);
   \draw[dotted,thick] (0.5,-1.091) -- (0.65,-1.35);
   \draw[dotted,thick] (0,-1.225) -- (0,-1.525);
   \draw[dotted,thick] (-0.5,-1.091) -- (-0.65,-1.35);
   \draw[dotted,thick] (-0.866,-0.725) -- (-1.126,-0.875);
   \draw[dashed,thick,blue] (1,-0.2) -- (-1,-0.2);
   \draw (0.4,-0.8) node{$A$};
   \draw (0.9,-1.4) node{$B$};
   \draw (-1.5,-0.2) node{$\beta$};
   \draw (1.5,-0.2) node{$0$};
\end{tikzpicture}.
\end{equation}
Here, the solid lines represent the imaginary-time evolution $e^{-\beta H}$. Quantum states are assigned to the open endpoints at $0$ and $\beta$, and specifying these boundary states yields the corresponding matrix element. The dotted lines represent interactions between the subsystems, while the red and blue dashed lines represent the Gaussian superoperators $T_A$ and $T_A^{-1}$, respectively. The moments $\mathcal{M}^{\bm{f}}_A$ are then obtained by gluing $2n$ density matrices in the order specified by $\bm{f}$. As an illustration, we present the corresponding contours for $2n=4$:
\begin{equation}
\begin{aligned}
\mathcal{M}^{(+,+,+,+)}_A=\frac{1}{Z^4}\times~
\begin{tikzpicture}[baseline={([yshift=0pt]current bounding box.center)}]
  \draw (1,-0.2) arc (0:-360:1 and 1);
  \draw (1.3,-0.2) arc (0:-360:1.3 and 1.3);
  \foreach \angle in {0,30,...,330} {
    \draw[dotted,thick,shift={(0,-0.2)}]
      (\angle:1) -- (\angle:1.3);
  }
  \draw[dashed,thick,red] (1,-0.2) -- (0,-1.2);
  \draw[dashed,thick,red] (1,-0.2) -- (0,0.8);
\draw[dashed,thick,red] (-1,-0.2) -- (0,-1.2);
  \draw[dashed,thick,red] (-1,-0.2) -- (0,0.8);

\fill[black] (1,-0.2) circle[radius=0.04];
\fill[black] (1.3,-0.2) circle[radius=0.04];
\end{tikzpicture}\ ,\ \ \ \ \ \ \ \ 
\mathcal{M}^{(+,+,+,-)}_A=\frac{1}{Z^4}\times~
\begin{tikzpicture}[baseline={([yshift=0pt]current bounding box.center)}]
  \draw (1,-0.2) arc (0:-360:1 and 1);
  \draw (1.3,-0.2) arc (0:-360:1.3 and 1.3);
  \foreach \angle in {0,30,...,330} {
    \draw[dotted,thick,shift={(0,-0.2)}]
      (\angle:1) -- (\angle:1.3);
  }
  \draw[dashed,thick,red] (1,-0.2) -- (0,-1.2);
  \draw[dashed,thick,blue] (1,-0.2) -- (0,0.8);
\draw[dashed,thick,red] (-1,-0.2) -- (0,-1.2);
  \draw[dashed,thick,red] (-1,-0.2) -- (0,0.8);

\fill[black] (1,-0.2) circle[radius=0.04];
\fill[black] (1.3,-0.2) circle[radius=0.04];
\end{tikzpicture}\ ,\\
\mathcal{M}^{(+,+,-,-)}_A=\frac{1}{Z^4}\times~
\begin{tikzpicture}[baseline={([yshift=0pt]current bounding box.center)}]
  \draw (1,-0.2) arc (0:-360:1 and 1);
  \draw (1.3,-0.2) arc (0:-360:1.3 and 1.3);
  \foreach \angle in {0,30,...,330} {
    \draw[dotted,thick,shift={(0,-0.2)}]
      (\angle:1) -- (\angle:1.3);
  }
  \draw[dashed,thick,red] (1,-0.2) -- (0,-1.2);
  \draw[dashed,thick,blue] (1,-0.2) -- (0,0.8);
\draw[dashed,thick,red] (-1,-0.2) -- (0,-1.2);
  \draw[dashed,thick,blue] (-1,-0.2) -- (0,0.8);

\fill[black] (1,-0.2) circle[radius=0.04];
\fill[black] (1.3,-0.2) circle[radius=0.04];
\end{tikzpicture}\ ,\ \ \ \ \ \ \ \ 
\mathcal{M}^{(+,-,+,-)}_A=\frac{1}{Z^4}\times~
\begin{tikzpicture}[baseline={([yshift=0pt]current bounding box.center)}]
  \draw (1,-0.2) arc (0:-360:1 and 1);
  \draw (1.3,-0.2) arc (0:-360:1.3 and 1.3);
  \foreach \angle in {0,30,...,330} {
    \draw[dotted,thick,shift={(0,-0.2)}]
      (\angle:1) -- (\angle:1.3);
  }
  \draw[dashed,thick,red] (1,-0.2) -- (0,-1.2);
  \draw[dashed,thick,blue] (1,-0.2) -- (0,0.8);
\draw[dashed,thick,blue] (-1,-0.2) -- (0,-1.2);
  \draw[dashed,thick,red] (-1,-0.2) -- (0,0.8);

\fill[black] (1,-0.2) circle[radius=0.04];
\fill[black] (1.3,-0.2) circle[radius=0.04];
\end{tikzpicture}\ .
\end{aligned}
\end{equation}
Here, the black dots indicate antiperiodic boundary conditions for the fermion fields. The resulting contour has a total imaginary-time extent of $2n\beta$. More explicitly, the moments $Z^{\bm{f}}_A\equiv Z^{2n}\mathcal{M}^{\bm{f}}_A $ admit the path-integral representation:
\begin{equation}
Z^{\bm{f}}_A=\int D\chi_j(\tau)\exp\Bigg(-\int_0^{2n\beta}d\tau~\Bigg[\frac{1}{2}\sum_j\chi_j\partial_\tau \chi_j+H[\chi_j]\Bigg]-\Delta S^{\bm{f}}[\chi_j]\Bigg).
\end{equation}
Here, $\Delta S^{\bm{f}}$ denotes the contribution from the Gaussian superoperators. This term is quadratic in the Majorana fields and couples $\chi_j((k+1)\beta)$ and $\chi_j(k\beta)$ for $j\in A$ and $k=0,1,\ldots,2n-1$. Although an explicit expression can be derived, we adopt an alternative perspective that is more convenient for subsequent calculations. The central observation is that, because $\Delta S^{\bm{f}}$ is quadratic, its effect can be incorporated into a modified free two-point function at $J=0$ for fermions in subsystem $A$. This leads to
\begin{equation}\label{eq:Zf}
Z^{\bm{f}}_A=\int D\chi_j(\tau)\exp\Bigg(-\int ~\Bigg[\frac{1}{2}\sum_{j\in A}\chi_j[G_A^{\bm{f},0}]^{-1}\chi_j+\frac{1}{2}\sum_{i\in B}\chi_i\partial_\tau \chi_i+H[\chi_j]\Bigg]\Bigg),
\end{equation}
where the first term represents a general quadratic term\footnote{The two Majorana fields generally have different time arguments, which we leave implicit.} with two-point function $G_A^{\bm{f},0}$. The free two-point function $G_A^{\bm{f},0}$ can be evaluated directly from its definition:
\begin{equation}
G_A^{\bm{f},0}(\tau,\tau')
=\langle \chi_j(\tau)\chi_j(\tau')\rangle_{J=0},
\qquad j\in A,
\end{equation}
There are two cases to consider. First, when $\tau$ and $\tau'$ lie within the same copy of the density matrix, accounting for fermionic contour ordering gives
\begin{equation}\label{eq:G1}
G_A^{\bm{f},0}(\tau,\tau')
=\frac{1}{2}\operatorname{sgn}(\tau-\tau'),
\qquad
\tau,\tau'\in(k\beta,(k+1)\beta).
\end{equation}
Second, when $\tau$ and $\tau'$ lie in different copies $a$ and $b$ of the density matrix, a straightforward application of the corresponding superoperators $T_A$ or $T_A^{-1}$ yields
\begin{equation}\label{eq:G2}
G_A^{\bm{f},0}(\tau,\tau')
=-\frac{1}{2}f_af_b\operatorname{sgn}(a-b),
\qquad a\neq b.
\end{equation}  
Combining Eqs.~\eqref{eq:G1} and \eqref{eq:G2} gives the full free Green's function for subsystem $A$ on a generic contour specified by $\bm{f}$. In contrast, the free Green's function for subsystem $B$ is given by $G_B^{\bm{f},0}(\tau,\tau')=\frac{1}{2}\operatorname{sgn}(\tau-\tau')$ for arbitrary $\tau$ and $\tau'$. Together, these Green's functions determine the path integral in Eq.~\eqref{eq:Zf}. This formulation may also be applied to Monte Carlo simulations of the generalized moments $\mathcal{M}^{\bm{f}}_A$. 

\subsection{The SYK Analysis}

To proceed, we need to employ specific properties of the SYK Hamiltonian. First, the SYK model is known to be diagonal in disorder replica space, implying that it is self-averaging~\cite{kitaev2018soft,Gu:2019jub}. As a result, when computing the disorder average of the relative R\'enyi negativities in the large-$N$ limit, we can safely approximate:
\begin{equation}
\overline{\mathcal{P}^{\bm{f}}_A}= -\text{Re}\left[\overline{\ln \left(\mathcal{M}^{\bm{f}}_A/ \mathcal{M}^{\bm{f}_t}_A\right)}\right]\approx -\text{Re}\left[\ln\left(\overline{Z_A^{\bm{f}}}/\overline{Z_A^{\bm{f}_t}}\right)\right].
\end{equation}
Then, the disorder-averaged partition function can be calculated directly using the Gaussian distribution of the couplings $J_{i_1i_2\cdots i_q}$:
\begin{equation} 
\overline{Z_A^{\bm{f}}}=\int dJ_{i_1i_2\cdots i_q} P(J_{i_1i_2\cdots i_q})\int D\chi_j(\tau)e^{-\int ~\left[\frac{1}{2}\sum_{j\in A}\chi_j[G_A^{\bm{f},0}]^{-1}\chi_j+\frac{1}{2}\sum_{i\in B}\chi_i\partial_\tau \chi_i+H[\chi_j]\right]}.
\end{equation}
Here, $P(J_{i_1i_2\cdots i_q})$ denotes the Gaussian distribution function. Following the standard SYK technique~\cite{maldacena2016remarks}, we integrate over the couplings $J_{i_1i_2\cdots i_q}$ and introduce the bilocal fields $G_{A/B}(\tau,\tau')=\sum_{j\in A/B}\chi_j(\tau)\chi_j(\tau')/N_{A/B}$ and the corresponding self-energy fields $\Sigma_{A/B}(\tau,\tau')$. This yields $\overline{Z_A^{\bm{f}}}=\int D\chi_j\,DG_{A/B}\,D\Sigma_{A/B}\,e^{-S[\chi_j,G_{A/B},\Sigma_{A/B}]}$, with
\begin{equation} 
\begin{aligned}
S=&\int_{\tau,\tau'}~\frac{1}{2}\sum_{j\in A}\chi_j(\tau)\Big((G_{A}^{\bm{f},0})^{-1}-\Sigma_{A}\Big)\chi_j(\tau')+\frac{1}{2}\sum_{j\in B}\chi_j(\tau)(\delta_{\tau\tau'}\partial_\tau-\Sigma_{B})\chi_j(\tau')\\
&+\sum_{A/B}\frac{N_{A/B}}{2}G_{A/B}(\tau,\tau')\Sigma_{A/B}(\tau,\tau')-\frac{J^2N}{2q}\left(\frac{N_A}{N}G_A(\tau,\tau')+\frac{N_B}{N}G_B(\tau,\tau')\right)^q.
\end{aligned}
\end{equation}
Next, we integrate out the fermionic modes to obtain the $G$-$\Sigma$ effective action $S_{G\Sigma}$:
\begin{equation}
\begin{aligned}\label{eq:GSigma}
S_{G\Sigma}&=-\frac{N_A}{2}\ln \det\left[(G_{A}^{\bm{f},0})^{-1}-\Sigma_{A}\right]-\frac{N_B}{2}\ln \det\left[\partial_\tau-\Sigma_{B}\right]\\&+\int_{\tau\tau'}\sum_{A/B}\frac{N_{A/B}}{2}G_{A/B}(\tau,\tau')\Sigma_{A/B}(\tau,\tau')-\frac{J^2N}{2q}\left(\frac{N_A}{N}G_A(\tau,\tau')+\frac{N_B}{N}G_B(\tau,\tau')\right)^q.
\end{aligned}
\end{equation}
The action is now explicitly proportional to $N$. In the large-$N$ limit, we can therefore apply the saddle-point approximation to the bilocal fields $G_{A/B}$ and $\Sigma_{A/B}$. The resulting saddle-point equations are given by
\begin{equation}\label{eq:saddle}
\begin{aligned}
&G_A=\Big((G_{A}^{\bm{f},0})^{-1}-\Sigma_{A}\Big)^{-1},\ \ \ \ \ \ G_B=\left(\partial_\tau-\Sigma_{B}\right)^{-1},\\
&\Sigma_A(\tau,\tau')=\Sigma_B(\tau,\tau')=J^2g(\tau,\tau')^{q-1}\equiv \Sigma(\tau,\tau').
\end{aligned}
\end{equation}
Here, we have introduced $g(\tau,\tau')=\frac{N_A}{N}G_A(\tau,\tau')+\frac{N_B}{N}G_B(\tau,\tau')$ for brevity. After solving the saddle-point equations, we obtain the partition function by evaluating the on-shell action $I_{G\Sigma}$. This gives
\begin{equation}
-\ln \overline{Z_A^{\bm{f}}}=I_{G\Sigma}=\frac{N_A}{2}\ln \det G_A+\frac{N_B}{2}\ln \det G_B+\frac{N(q-1)}{2q}\int_{\tau,\tau'}g(\tau,\tau')\Sigma(\tau,\tau'),
\end{equation}
where we have used the saddle-point equations. This determines the extensive contribution to the relative R\'enyi negativities in the thermodynamic limit $N\rightarrow\infty$.

  \begin{figure}[t]
    \centering
    \includegraphics[width=1.0\linewidth]{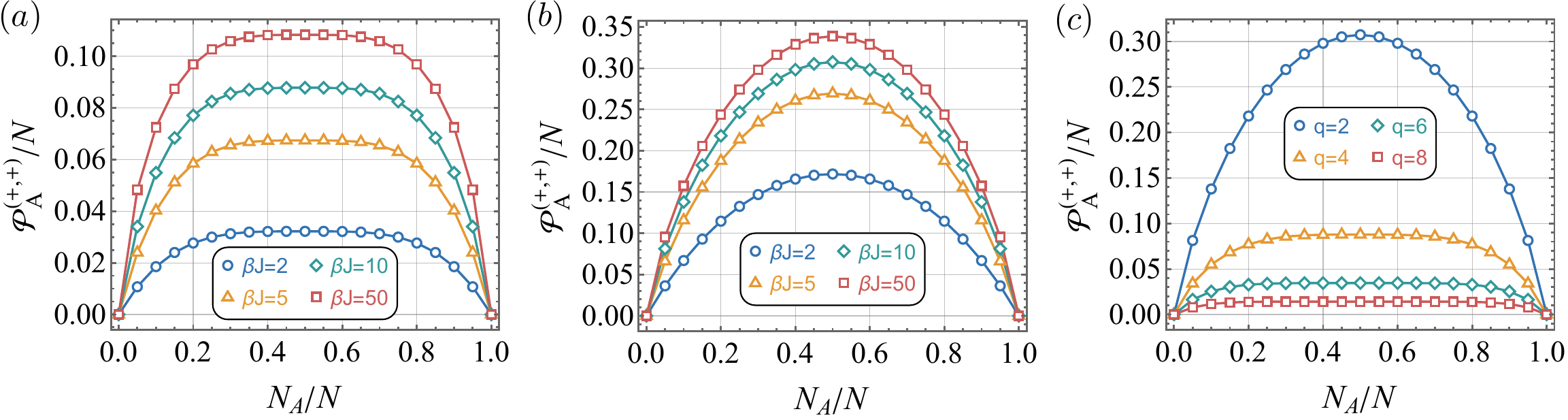}
    \caption{Numerical results for the relative R\'enyi negativity $\mathcal{P}^{(+,+)}_A$. Panels (a) and (b) show the results as a function of the subsystem fraction $N_A/N$ for $q=4$ and $q=2$, respectively, with various $\beta J\in\{2,5,10,50\}$. Panel (c) compares the results for different values of $q$ at fixed $\beta J=10$.}
    \label{fig:num1}
  \end{figure}

Since the insertions of the superoperators $T_A$ and $T_A^{-1}$ break translation symmetry in imaginary time, solving the saddle-point equations analytically is challenging. We therefore solve them numerically by discretizing the imaginary-time interval $\tau\in[0,2n\beta)$ on a grid and representing the bilocal fields as matrices, following an approach similar to that in Ref.~\cite{Zhang:2020kia}. The saddle-point equations~\eqref{eq:saddle} then become matrix equations, which we solve iteratively. The partition function is evaluated using the converged solutions, and is always real in our numerics. 

We first focus on $2n=2$, where the only nontrivial relative R\'enyi negativity is $\mathcal{P}^{(+,+)}_A$. The numerical results for various values of $q$ and inverse temperature $\beta$ are presented in Figure~\ref{fig:num1} as functions of the subsystem fraction $N_A/N$. First, for both the chaotic model with $q=4$ and the free-fermion model with $q=2$, $\mathcal{P}^{(+,+)}_A$ decreases with increasing temperature. Higher temperatures increase the thermal entropy, which can be interpreted, through a purification of the thermal state, as greater entanglement between the entire system and an environment. Our results are consistent with the intuitive picture that thermal mixing suppresses entanglement between subsystems $A$ and $B$~\cite{Wang:2025mdu}. In particular, in the infinite-temperature limit, the density matrix becomes maximally mixed and $\mathcal{P}^{(+,+)}_A$ vanishes. For $q=4$, $\mathcal{P}^{(+,+)}_A$ exhibits a plateau near $N_A/N\approx1/2$, in qualitative agreement with previous results for the entanglement negativity of the reduced density matrix of a Haar-random state~\cite{PRXQuantum.2.030347}. In contrast, in the low-temperature limit, the negativity for $q=2$ exhibits a peak similar to that of the subsystem R\'enyi entropy, which has been determined analytically~\cite{Zhang:2020kia}. Moreover, $\mathcal{P}^{(+,+)}_A$ is substantially larger for $q=2$ than for $q=4$. This difference is qualitatively consistent with the presence of an extensive residual thermal entropy for $q=4$, given by $S_0/N=\int_{1/q}^{1/2}\pi(1/2-x)\tan(\pi x)\,dx$, which persists as $\beta J\to\infty$. This residual thermal mixedness can suppress entanglement within the system. Furthermore, $S_0/N$ increases toward its maximal value $\ln2/2$ as $q$ increases, consistent with the observed decrease in $\mathcal{P}^{(+,+)}_A$ in Figure~\ref{fig:num1}(c).

  \begin{figure}[t]
    \centering
    \includegraphics[width=1.0\linewidth]{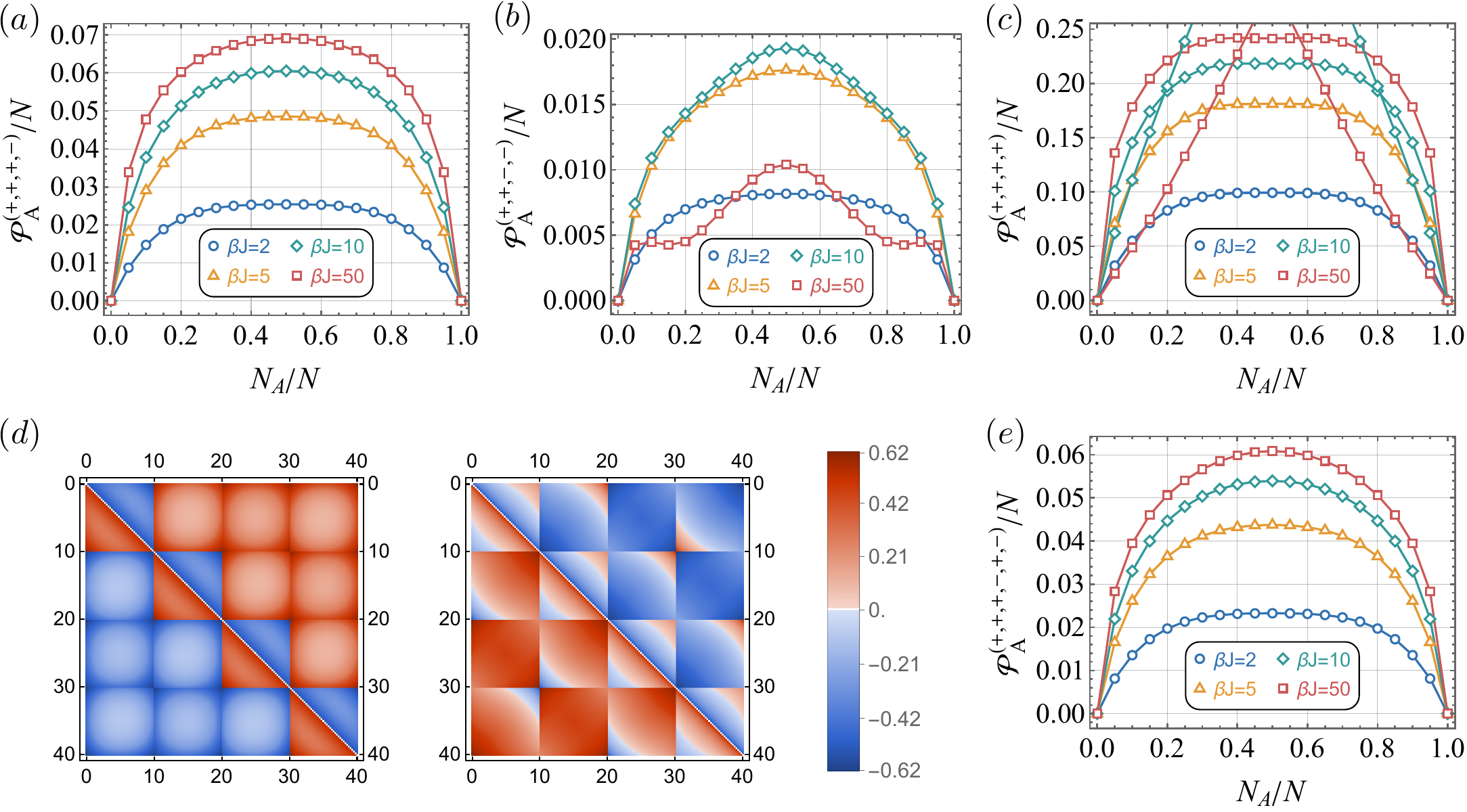}
    \caption{Numerical results for higher-order relative R\'enyi negativities at $q=4$ and $\beta J\in\{2,5,10,50\}$. Panels (a-c) show the three nontrivial relative R\'enyi negativities at fourth order, $2n=4$. Among these, only the minimal-deviation case, $\mathcal{P}^{(+,+,+,-)}_A$, decreases monotonically with increasing temperature for all subsystem fractions $N_A/N$. For $\mathcal{P}^{(+,+,+,+)}_A$, multiple saddle-point solutions coexist at low temperatures. Panel (d) displays the Green's functions $G_A(\tau,\tau')$ for two distinct saddles at $N_A/N=0.2$ and $\beta J=10$. The solution on the right dominates at small subsystem fractions $N_A/N$. Panel (e) shows the minimal-deviation case for $2n=6$.  }
    \label{fig:num2}
  \end{figure}

We next consider higher R\'enyi orders. At $2n=4$, there are three distinct nontrivial relative R\'enyi negativities. To investigate their temperature dependence, we fix $q=4$ and plot them as functions of $N_A/N$ at different temperatures. The results are shown in Fig.~\ref{fig:num2}(a--c). Although positive values of all three quantities certify quantum entanglement, they exhibit qualitatively different behavior. The minimal-deviation case, $\mathcal{P}^{\bm{f}_m}_A=\mathcal{P}^{(+,+,+,-)}_A$, behaves similarly to $\mathcal{P}^{(+,+)}_A$, decreasing monotonically with increasing temperature, consistent with the physical expectation for an entanglement measure. For $\mathcal{P}^{(+,+,-,-)}_A$, however, the negativity first increases and then decreases with increasing temperature, exhibiting nonmonotonic behavior. Nevertheless, it remains a smooth function of the subsystem fraction $N_A/N$. Finally, for $\mathcal{P}^{\bm{f}_u}_A=\mathcal{P}^{(+,+,+,+)}_A$, multiple saddle-point solutions for $Z^{\bm{f}_u}_A$ coexist at $\beta J\in\{10,50\}$. The physical result is determined by the saddle with the lowest on-shell action, which yields the smallest relative negativity. The exchange of dominance between these saddles leads to first-order transitions, with the relative negativity playing the role of an effective free energy. The resulting $\mathcal{P}^{\bm{f}_u}_A$ grows linearly at small $N_A/N$ before transitioning to a plateau near $N_A/N=1/2$. We emphasize that this transition is intrinsic to the negativity and is invisible to simple observables.

To understand these two saddle-point solutions, we plot the corresponding Green's functions $G_A(\tau,\tau')$ in panel (d). The saddle shown on the left dominates near $N_A/N=1/2$. Its sign structure resembles that of $G_A^{\bm{f}_u,0}$ (see Eq.~\eqref{eq:G2}), suggesting that interactions preserve the qualitative pattern of correlations between replicas. By contrast, the saddle shown on the right dominates at small $N_A/N$. Its inter-replica correlations are predominantly positive for $\tau>\tau'$, resembling the structure of $G_B^0$. In this regime, the effect of $T_A$ can be viewed as a perturbation. This interpretation is consistent with the suppression of correlations between Majorana fields separated by imaginary-time intervals of order $\beta$. It provides a qualitative explanation for the observed decrease in $\mathcal{P}^{\bm{f}_u}_A$ with increasing $\beta J$ in the low-temperature regime, which leads to the nonmonotonic temperature dependence at small $N_A/N$. The presence of different saddle-point solutions is reminiscent of replica-wormhole saddles observed in the dynamics of the second R\'enyi entropy in the SYK model~\cite{almheiri2020replica,Penington:2019kki,chen2020replica,zhang2020entanglement,PhysRevD.109.046005}.

Taken together, these results show that $\mathcal{P}^{(+,+)}_A$ and $\mathcal{P}^{(+,+,+,-)}_A$ exhibit behavior consistent with physical expectations. Both belong to the class $\bm{f}=\bm{f}_m$, corresponding to a minimal deviation from the alternating sign sequence $\bm{f}_t$. To further assess whether members of this class exhibit desirable behavior as proxies for logarithmic entanglement negativity, we compute the sixth-order quantity $\mathcal{P}^{(+,+,+,-,+,-)}_A$. The result, presented in Fig.~\ref{fig:num2}(e), again shows a monotonic decrease with increasing temperature, supporting this expectation. In the following discussion, we therefore focus on $\mathcal{P}^{(+,+)}_A$ and $\mathcal{P}^{(+,+,+,-)}_A$ to investigate mixed-state entanglement in a generalized SYK model that exhibits a thermal phase transition.

\section{Negativities in the Coupled SYK Model}
In this section, we apply the framework established in the previous section to study entanglement in the thermal ensemble of the Maldacena-Qi coupled SYK model. Following the convention in Ref.~\cite{Maldacena:2018lmt}, the entire system contains $2N$ Majorana fermions, divided into two subsystems: left ($L$) and right ($R$). The corresponding Majorana operators are denoted by $\chi_{j,\eta}$, with $j\in\{1,2,\ldots,N\}$ and $\eta\in\{L,R\}$. The Hamiltonian reads~\cite{Maldacena:2018lmt}
\begin{equation}
H_{MQ}=\sum_\eta\sum_{j<k<l<m}J_{jklm}~\chi_{j,\eta}\chi_{k,\eta}\chi_{l,\eta}\chi_{m,\eta}+\sum_j i\mu \chi_{j,L}\chi_{j,R}.
\end{equation}
Here, the first term describes the SYK interactions within each subsystem, with identical random couplings $J_{jklm}$ on both sides. The second term describes a direct coupling between the two sides, which induces entanglement between the subsystems. In the weakly coupled regime, $\mu/J\ll1$, the system exhibits a first-order thermal transition as the dimensionless inverse temperature $\beta J$ is varied, reminiscent of the Hawking-Page transition. The low-temperature phase has an emergent description as an eternal traversable wormhole in approximately global $\mathrm{AdS}_2$ spacetime, with strong entanglement between the two sides encoded geometrically. In the high-temperature phase, the coupling between the two sides can be treated perturbatively, and the geometric picture consists of two nearly independent black holes with weak quantum correlations between them.

We investigate the relative R\'enyi negativities between the two sides of the coupled SYK model by adapting the framework established in the previous section. Setting $A=L$, the corresponding partition function specified by $\bm{f}$ is given by
\begin{equation}
\begin{aligned}
Z^{\bm{f}}_L&=\int D\chi_j(\tau)\exp\left(-\int ~\bigg[\frac{1}{2}\sum_{j}\chi_{j,L}[G_L^{\bm{f},0}]^{-1}\chi_{j,L}+\frac{1}{2}\sum_{j}\chi_{j,R}\partial_\tau\chi_{j,R}+H[\chi_{j,\eta}]\bigg]\right).\\
\end{aligned}
\end{equation}
After performing the disorder average and introducing the bilocal fields, we obtain the coupled-SYK analogue of the $G$-$\Sigma$ action in Eq.~\eqref{eq:GSigma}: 
\begin{equation}
\begin{aligned}
S_{G\Sigma}
={}&-\frac{N}{2}\ln\det
\begin{pmatrix}
(G_L^{\bm{f},0})^{-1}-\Sigma_{LL}
& i\mu-\Sigma_{LR}\\
-i\mu-\Sigma_{RL}
& \partial_\tau-\Sigma_{RR}
\end{pmatrix}\\
&+\frac{N}{2}\int_{\tau\tau'}
\sum_{\eta,\eta'\in\{L,R\}}
\left[
G_{\eta\eta'}(\tau,\tau')\Sigma_{\eta\eta'}(\tau,\tau')
-\frac{J^2}{4}G_{\eta\eta'}(\tau,\tau')^4
\right].
\end{aligned}
\end{equation}
Here, the free Green's function $G_L^{\bm{f},0}$ is again given by Eqs.~\eqref{eq:G1} and \eqref{eq:G2}. In the large-$N$ limit, varying this action yields the saddle-point equations
\begin{equation}
\begin{aligned}
G\equiv \begin{pmatrix}
G_{LL} & G_{LR}\\
G_{RL} & G_{RR}
\end{pmatrix}
&=
\begin{pmatrix}
(G_L^{\bm{f},0})^{-1}-\Sigma_{LL}
& i\mu-\Sigma_{LR}\\
-i\mu-\Sigma_{RL}
& \partial_\tau-\Sigma_{RR}
\end{pmatrix}^{-1},\ \ \ \ 
\Sigma_{\eta\eta'}(\tau,\tau')
&=J^2G_{\eta\eta'}(\tau,\tau')^3.
\end{aligned}
\label{eq:MQ-saddle}
\end{equation}
We have assembled the Green's functions into a $2\times2$ block matrix and the inverse acts on both subsystem and imaginary-time indices. After solving the saddle-point equations, the disorder-averaged partition function $\overline{Z_L^{\bm{f}}}$ is given by the on-shell action 
\begin{equation}
-\ln \overline{Z_L^{\bm{f}}}=\frac{N}{2}\ln \det G+\frac{3N}{8}\int_{\tau\tau'}
\sum_{\eta,\eta'\in\{L,R\}}
G_{\eta\eta'}(\tau,\tau')\Sigma_{\eta\eta'}(\tau,\tau').
\end{equation}

  \begin{figure}[t]
    \centering
    \includegraphics[width=0.8\linewidth]{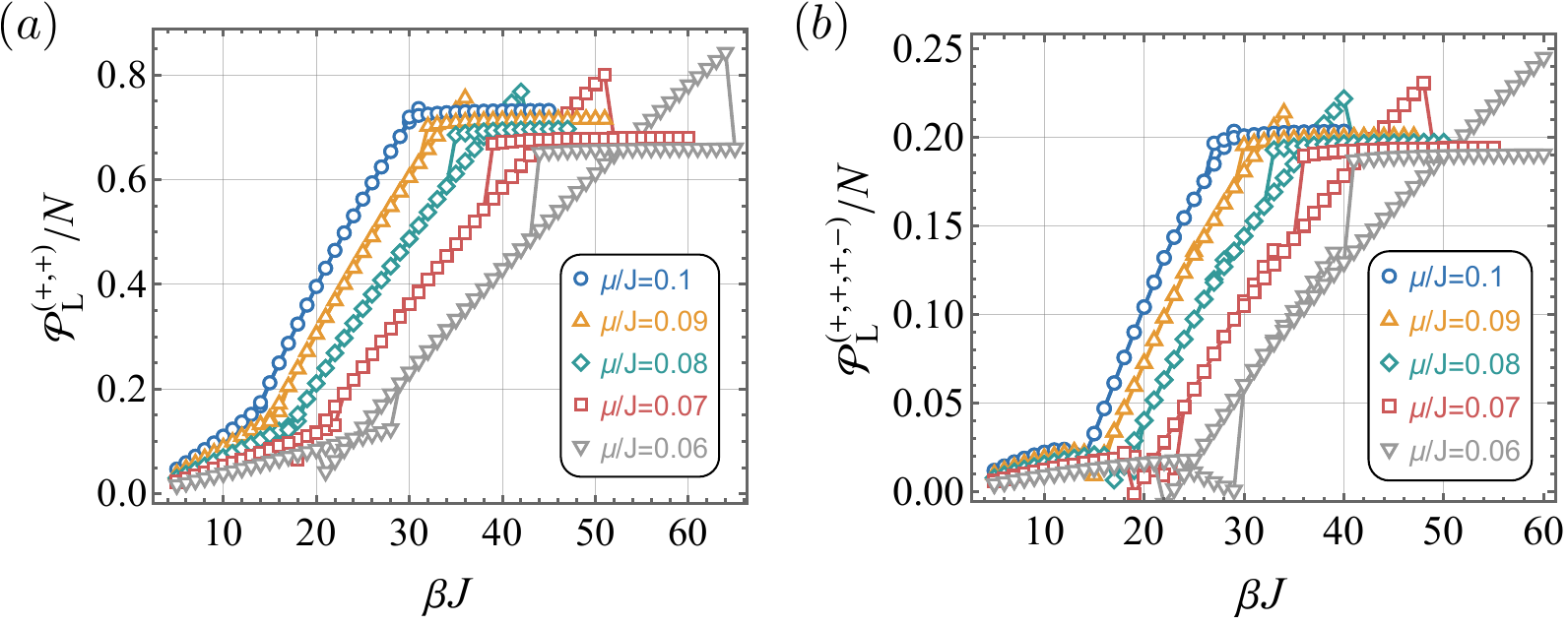}
    \caption{Numerical results for the relative R\'enyi negativities $\mathcal{P}^{\bm{f}_m}_L=-\text{Re}\left[\ln\left(\overline{Z_L^{\bm{f}_m}}/\overline{Z_L^{\bm{f}_t}}\right)\right]$ at orders $2n=2,4$ in the coupled SYK model with various values of $\mu/J$ in the weakly coupled regime. Both negativities exhibit two first-order transitions, arising from changes in the dominant saddle of either $\overline{Z_L^{\bm{f}_m}}$ or $\overline{Z_L^{\bm{f}_t}}$. Consequently, the physical solution corresponds to the branch with larger $\mathcal{P}_L^{\bm{f}_m}$ near the transition at smaller $\beta J$ and the branch with smaller $\mathcal{P}_L^{\bm{f}_m}$ near the transition at larger $\beta J$. }
    \label{fig:num3}
  \end{figure}

The numerical results for $\bm{f}=(+,+)$ and $\bm{f}=(+,+,+,-)$ are presented in Fig.~\ref{fig:num3}, where we plot the relative R\'enyi negativities as functions of the dimensionless inverse temperature $\beta J$ for various values of $\mu/J$ in the weakly coupled regime. Both relative R\'enyi negativities exhibit two successive first-order transitions as the temperature decreases. For $\mathcal{P}_L^{(+,+)}$, shown in panel (a), we have $\mathcal{P}_L^{(+,+)}=-\ln\left[\overline{Z_L^{(+,+)}}/Z(2\beta)\right]$. Consequently, the transition at smaller $\beta J$ originates from the thermal phase transition in $Z(2\beta)$, and the physical solution corresponds to the branch with the larger value of $\mathcal{P}_L^{(+,+)}$. For $\mu/J=0.1$, this transition occurs at $\beta J\approx14$, consistent with the thermal transition reported in Ref.~\cite{Maldacena:2018lmt} after accounting for the doubled inverse temperature. The transition at larger $\beta J$ originates from $\overline{Z_L^{(+,+)}}$ and is intrinsic to the relative R\'enyi negativity. Here, the physical solution corresponds to the branch with the smaller value of $\mathcal{P}_L^{(+,+)}$. Interestingly, although the contour contains $2n=2$ replicas, this transition occurs at approximately twice the inverse temperature of the first transition and therefore lies close to the thermal transition of the single-replica system. This suggests that correlations between different replicas become negligible in the low-temperature phase. Such a qualitative change in inter-replica correlations resembles that observed across the transition in the single-SYK model, although here it is driven by the coupling between the two sides.

Finally, the same qualitative features persist for $\bm{f}=(+,+,+,-)$, although additional subdominant saddles exist at intermediate temperatures. This suggests that, in the low-temperature phase, inter-replica correlations in $Z_L^{(+,-,+,-)}$ organize the replicas into nearly independent pairs, effectively dividing the contour into two-replica segments, whereas inter-replica correlations in $Z_L^{(+,+,+,-)}$ become negligible. These results illustrate how the rich replica structures underlying this zoo of relative R\'enyi negativities can reveal entanglement transitions beyond those captured by conventional thermodynamic observables.

\section{Discussions}
In this work, we introduced a zoo of relative R\'enyi negativities for fermionic many-body systems by comparing different orderings of the partially transposed density matrix and its Hermitian conjugate. These quantities are non-negative and vanish for separable states, so their positive values certify mixed-state entanglement. By expressing partial time reversal through the size superoperator, we developed a path-integral representation at fixed integer replica number and applied it to the original SYK model and the Maldacena-Qi coupled SYK model in the large-$N$ limit. Our results reveal distinct behavior among members of this family. In particular, the minimal-deviation sequences exhibit a monotonic decrease with increasing temperature in the single-SYK model, making them promising proxies for logarithmic negativity. Competing replica saddles also produce first-order transitions in the relative negativities. In the coupled model, two successive transitions arise from changes in the dominant saddles of the reference and numerator partition functions, respectively. These findings demonstrate that relative R\'enyi negativities can reveal entanglement structure beyond that captured by conventional thermodynamic observables.

We conclude by outlining several directions for future research. First, analytical studies in the large-$q$ or low-energy limits could clarify the structure of the competing replica saddles and the mechanisms governing their exchange of dominance. Such studies may also provide a gravitational interpretation of the transitions observed in the relative R\'enyi negativities. Second, it would be interesting to develop a perturbative framework for relative R\'enyi negativities based on the size superoperator. This approach may connect to recent developments in scramblon effective theory, which captures the universal exponential growth of operator size in systems with all-to-all interactions~\cite{gu2022two,Stanford:2021bhl}. Finally, extending the path-integral construction to numerical simulations of spatially local fermionic models and to nonequilibrium settings would test the broader applicability of these entanglement witnesses and reveal how the associated transitions depend on spatial structure and dynamical evolution.

\section*{Acknowledgments}
We thank Langxuan Chen, Zeyu Liu, Ning Sun, and Hanteng Wang for helpful discussions. This project is supported by the NSFC under grant 12374477, the Shanghai Rising-Star Program under grant number 24QA2700300, the Quantum Science and Technology-National Science and Technology Major Project 2024ZD0300101, the Shanghai Qiguang Natural Science Development Foundation, and the Xuemin Institute of Advanced Studies at Fudan University.

\appendix

\section{A separable state with positive R\'enyi negativity ratios}
\label{example1}

We construct an explicit example using four Majorana fermions in each subsystem, so that $N_A=N_B=4$. For later convenience, we write $\chi_{A,j}=\chi_j$ and $\chi_{B,j}=\chi_{4+j}$ for $j=1,\ldots,4$. The corresponding complex fermions are
\begin{equation}
f_{\alpha,1}=\frac{\chi_{\alpha,1}+i\chi_{\alpha,2}}{\sqrt{2}},
\qquad \ \ \ 
f_{\alpha,2}=\frac{\chi_{\alpha,3}+i\chi_{\alpha,4}}{\sqrt{2}}.
\end{equation}
We define the Fock state $|11\rangle_\alpha=f_{\alpha,1}^\dagger f_{\alpha,2}^\dagger|00\rangle_\alpha$ and encode a local qubit in the even-parity sector as $|0\rangle_\alpha\equiv|00\rangle_\alpha$ and $|1\rangle_\alpha\equiv|11\rangle_\alpha$. Consider the state
\begin{equation}
\rho_{\mathrm{sep}}=\frac{1}{3}\left(
|0,0\rangle\langle0,0|
+|+,+\rangle\langle+,+|
+|+i,-i\rangle\langle+i,-i|
\right).
\label{eq:sep-state}
\end{equation}
Here, $|\pm\rangle=(|0\rangle\pm|1\rangle)/\sqrt{2}$
and $|\pm i\rangle=(|0\rangle\pm i|1\rangle)/\sqrt{2}$. This state is separable across $A|B$, since each term is a product of local states with definite even fermion parity.

To express the density matrix in terms of Majorana operators, we abbreviate an ordered string as $\chi_{j_1\cdots j_m}\equiv\chi_{j_1}\cdots\chi_{j_m}$ and define
\begin{equation}
D_A=\mathbb{I}-4\chi_{1234},
\qquad \ \ \ \ 
D_B=\mathbb{I}-4\chi_{5678}.
\end{equation}
Each $D_\alpha$ is twice the projector onto the local even-parity sector. The relevant local projectors have the expansions
\begin{equation}
\begin{aligned}
|0\rangle_A\langle0|
&=\frac{1}{4}\left[D_A-2i(\chi_{12}+\chi_{34})\right],\\
|+\rangle_A\langle+|
&=\frac{1}{4}\left[D_A-2i(\chi_{14}+\chi_{23})\right],\\
|\pm i\rangle_A\langle\pm i|
&=\frac{1}{4}\left[D_A\pm2i(\chi_{13}-\chi_{24})\right].
\end{aligned}
\end{equation}
The corresponding expressions in $B$ follow by shifting every Majorana index by four. Substituting these expansions into Eq.~\eqref{eq:sep-state} gives
\begin{equation}
\begin{aligned}
\rho_{\mathrm{sep}}=\frac{1}{48}\bigl\{
&[D_A-2i(\chi_{12}+\chi_{34})]\times[D_B-2i(\chi_{56}+\chi_{78})]\\
+{}&[D_A-2i(\chi_{14}+\chi_{23})]\times[D_B-2i(\chi_{58}+\chi_{67})]\\
+{}&[D_A+2i(\chi_{13}-\chi_{24})]\times[D_B-2i(\chi_{57}-\chi_{68})]\bigr\}.
\end{aligned}
\label{eq:sep-majorana-density}
\end{equation}
We now apply partial time reversal directly to this expansion. Under $T_A$, a Majorana monomial containing $k_A$ operators in $A$ acquires the phase $i^{k_A}$. Thus, partial time reversal reverses the signs of the bilinears in each $A$ factor, yielding
\begin{equation}
\begin{aligned}
\rho_{\mathrm{sep}}^{T_A}=\frac{1}{48}\bigl\{
&[D_A+2i(\chi_{12}+\chi_{34})]\times[D_B-2i(\chi_{56}+\chi_{78})]\\
+{}&[D_A+2i(\chi_{14}+\chi_{23})]\times[D_B-2i(\chi_{58}+\chi_{67})]\\
+{}&[D_A-2i(\chi_{13}-\chi_{24})]\times[D_B-2i(\chi_{57}-\chi_{68})]\bigr\}.
\end{aligned}
\label{eq:sep-majorana-ptr}
\end{equation}
Equivalently, the transformed operator can be written as
\begin{equation}
\rho_{\mathrm{sep}}^{T_A}=\frac{1}{3}\left(
|1,0\rangle\langle1,0|
+|-,+\rangle\langle-,+|
+|-i,-i\rangle\langle-i,-i|
\right).
\end{equation}
It is therefore Hermitian, positive semidefinite, and has unit trace. Consequently, its trace norm is unity and the logarithmic negativity vanishes, $\mathcal{N}_A=\ln\|\rho_{\mathrm{sep}}^{T_A}\|_1=0$. However, a direct evaluation gives
\begin{equation}
\begin{aligned}
\operatorname{tr}(\rho_{\mathrm{sep}}^4)
=\frac{43}{216},\ \ \ \ \ \ \ \ \ 
\operatorname{tr}\left[(\rho_{\mathrm{sep}}^{T_A})^4\right]
=\operatorname{tr}\left[
\left((\rho_{\mathrm{sep}}^{T_A})^\dagger
\rho_{\mathrm{sep}}^{T_A}\right)^2
\right]
=\frac{35}{216}.
\end{aligned}
\end{equation}
The untwisted and twisted fourth moments coincide because $\rho_{\mathrm{sep}}^{T_A}$ is Hermitian. Hence, the corresponding R\'enyi negativity ratios satisfy
\begin{equation}
\mathcal{R}_A^{(4)}
=\tilde{\mathcal{R}}_A^{(4)}
=\frac{1}{3}\ln\frac{43}{35}>0.
\end{equation}
This explicitly demonstrates that subtracting the global R\'enyi entropy does not make a positive untwisted or twisted R\'enyi negativity ratio a sufficient criterion for entanglement.

\section{Proof of the relation \texorpdfstring{$\mathcal{P}^{\bm{f}}_A=\mathcal{P}^{\bm{f}}_B$}{P_A(f) = P_B(f)}}
\label{proof}

In this appendix, we prove that the relative R\'enyi negativities are invariant under exchanging subsystems $A$ and $B$. The proof applies to any density matrix $\rho$ that conserves total fermion parity. We first consider the full fermionic transpose $T=T_A T_B$. Since $\rho$ preserves fermion parity, only Majorana strings with even total size $s=|\bm{a}|+|\bm{b}|$ contribute to Eq.~\eqref{eq:rho}. On such strings, the full transpose acts as
\begin{equation}
\begin{aligned}
T[\chi_{j_1}\chi_{j_2}\cdots\chi_{j_s}]
&=i^s\chi_{j_1}\chi_{j_2}\cdots\chi_{j_s}=(-1)^{s(s-1)/2}
\chi_{j_1}\chi_{j_2}\cdots\chi_{j_s}=\chi_{j_s}\cdots\chi_{j_2}\chi_{j_1}.
\end{aligned}
\end{equation}
Thus, the linear map $T$ reverses the order of the Majorana operators. It follows that, for operators that conserve total fermion parity,
\begin{equation}
T[O_1O_2]=T[O_2]T[O_1],
\qquad
T[O^\dagger]=(T[O])^\dagger,
\qquad
\operatorname{tr}T[O]=\operatorname{tr}O.
\label{eq:full-transpose-identities}
\end{equation}
Using $T=T_A T_B$ and the commutativity of $T_A$ and $T_B$, we find
\begin{equation}
\rho^{T_B}=T[T_A^{-1}[\rho]]=T[(\rho^{T_A})^\dagger],
\qquad
(\rho^{T_B})^\dagger=T[\rho^{T_A}].
\label{eq:complementary-partial-transpose}
\end{equation}
Therefore, applying Eq.~\eqref{eq:full-transpose-identities} gives
\begin{equation}
\begin{aligned}
\mathcal{M}_B^{\bm{f}}
&=\operatorname{tr}\left[
\prod_{j=1}^{2n}
\left(
\frac{1+f_j}{2}\rho^{T_B}
+\frac{1-f_j}{2}(\rho^{T_B})^\dagger
\right)
\right]\\
&=\operatorname{tr}T\left[
\left\{
\prod_{j=1}^{2n}
\left(
\frac{1+f_j}{2}\rho^{T_A}
+\frac{1-f_j}{2}(\rho^{T_A})^\dagger
\right)
\right\}^{\dagger}
\right]=\left(\mathcal{M}_A^{\bm{f}}\right)^*.
\end{aligned}
\label{eq:moment-subsystem-symmetry}
\end{equation}
Here, the products are ordered with increasing $j$ from left to right. The second equality follows because both the full transpose and Hermitian conjugation reverse the order of multiplication. Consequently,
\begin{equation}
\left|\mathcal{M}_A^{\bm{f}}\right|
=\left|\mathcal{M}_B^{\bm{f}}\right|.
\end{equation}
For the alternating reference sequence $\bm{f}_t$, the moments
\begin{equation}
\mathcal{M}_A^{\bm{f}_t}
=\operatorname{tr}\left[
\left((\rho^{T_A})^\dagger\rho^{T_A}\right)^n
\right],
\qquad
\mathcal{M}_B^{\bm{f}_t}
=\operatorname{tr}\left[
\left((\rho^{T_B})^\dagger\rho^{T_B}\right)^n
\right]
\end{equation}
are positive. Equation~\eqref{eq:moment-subsystem-symmetry} therefore implies
$\mathcal{M}_A^{\bm{f}_t}=\mathcal{M}_B^{\bm{f}_t}$.
Combining these results yields
\begin{equation}
\mathcal{P}_A^{\bm{f}}
=-\ln\left[
|\mathcal{M}_A^{\bm{f}}|/
{\mathcal{M}_A^{\bm{f}_t}}
\right]
=-\ln\left[
|\mathcal{M}_B^{\bm{f}}|/
{\mathcal{M}_B^{\bm{f}_t}}
\right]
=\mathcal{P}_B^{\bm{f}}.
\end{equation}
This establishes the desired symmetry for arbitrary $\bm{f}$.

\bibliography{Draft.bbl}

\end{document}